\documentclass[12pt,a4paper]{article}
\usepackage[T1]{fontenc}
\usepackage[sc,osf]{mathpazo}
\usepackage{a4wide}  
\usepackage{amsfonts}
\usepackage{amssymb}
\usepackage{amsmath}
\usepackage{bbm}
\usepackage{booktabs} 
\usepackage{ifpdf}
\ifpdf
\usepackage[pdftex,unicode,implicit]{hyperref}
\hypersetup{%
  pdftitle    = {Resolution of SU(2) monopole singularities by oxidation}
  pdfkeywords = {Yang-Mills, non-Abelian, monopole, singular monopole, colored,
instanton, Wu-Yang, Protogenov, Belavin-Polyakov-Schwarz-Tyupkin instanton, 
Higgs,  't~Hooft-Polyakov monopole, Bogomol'nyi},
  pdfauthor   = {Pablo Bueno, Pedro Fernandez-Ramirez, Patrick Meessen and Tomas Ortin},
  plainpages  = true,
  colorlinks  = true,
  citecolor   = blue,
  urlcolor    = red,
  linkcolor   = black
}
\newcommand{\hepth}[1]{{\tt
\href{http://www.arXiv.org/abs/hep-th/#1}{hep-th/#1}}}

\newcommand{\arxiv}[1]{{\tt
\href{http://www.arXiv.org/abs/#1}{#1}}}
\else
  \usepackage[dvips]{graphicx}
  \usepackage[unicode,implicit]{hyperref}
  \newcommand{\hepth}[1]{{\tt hep-th/#1}}

  \newcommand{\arxiv}[1]{{\tt arXiv:#1}}
\fi
\usepackage{tikz}\newcommand{\FPAUO}[2]{
\tikz[scale=.13,
         Uniovi/.style={color=green!51!blue, fill=green!51!blue}
 ] {
 \fill[Uniovi] (0,0) circle (10);
 \fill[white] (0,7) circle (1.5);
 \draw[Uniovi] (-2,7.5) rectangle (2,5.5);
 \fill[white] (-0.3,6.6) rectangle (0.3,0);   
 \fill[white] ( -0.9,6.2) rectangle (.9 ,5.6);
 \fill[white] (-1.4, 5.2) rectangle (1.4, 4.6);
 \fill[white] (0,0) ellipse (3.5 and 4);
 \fill[Uniovi] (-2.5,0.3) rectangle (2.5,-0.3);
 \fill[Uniovi] (-2,2.3) rectangle (2,1.7);
 \fill[Uniovi] (-2,-2.3) rectangle (2,-1.7);
 \fill[white] (-4.5,5.5) rectangle (-2.7,4.9);
 \fill[white] (-3.9,6.1) rectangle (-3.3,4.3);
 \fill[white] (4.5,5.5) rectangle (2.7,4.9);
 \fill[white] (3.9,6.1) rectangle (3.3,4.3);
 \foreach \x in { 0,..., 3 }
   \foreach \y in { 0,...,\x}
    {
     \fill[white] (-6-\x*0.7+\y*1.4,3.5-\x *1.97) -- (-5.6-\x*0.7+\y*1.4,2.4-\x *1.97) -- (-6.4-\x*0.7+\y*1.4,2.4-\x *1.97) -- cycle;
     \fill[white] (6-\x*0.7+\y*1.4,3.5-\x *1.97) -- (5.6-\x*0.7+\y*1.4,2.4-\x *1.97) -- (6.4-\x*0.7+\y*1.4,2.4-\x *1.97) -- cycle;
   };
 \draw (0,-6) node[
                               text centered, 
                               color=white, 
                               font={\fontsize{8}{4}\sffamily\selectfont}
                             ] {FPAUO-#1/#2};
}} 
\usepackage{pgfplots}    
\makeatletter
\@addtoreset{equation}{section}
\makeatother

\begin{document}

\begin{flushright}
\small
\FPAUO{15}{07}\\
IFT-UAM/CSIC-15-019\\
\texttt{arXiv:yymm.nnnn [hep-th]}\\
March 3\textsuperscript{rd}, 2015\\
\normalsize
\end{flushright}

\vspace{1.5cm}

\begin{center}

{\Large {\bf Resolution of $\mathrm{SU}(2)$ monopole singularities by oxidation}}

\vspace{1.5cm}

\renewcommand{\thefootnote}{\alph{footnote}}
{\sl\large  Pablo Bueno$^{1}$}${}^{,}$\footnote{E-mail: {\tt p.bueno [at] csic.es}},
{\sl\large Patrick Meessen$^{2}$}${}^{,}$\footnote{E-mail: {\tt meessenpatrick [at] uniovi.es}},
{\sl\large Tom\'{a}s Ort\'{\i}n$^{1}$}${}^{,}$\footnote{E-mail: {\tt Tomas.Ortin [at] csic.es}}
{\sl\large and Pedro F.~Ram\'{\i}rez$^{1}$}${}^{,}$\footnote{E-mail: {\tt p.f.ramirez [at]  csic.es}},

\setcounter{footnote}{0}
\renewcommand{\thefootnote}{\arabic{footnote}}

\vspace{1.5cm}

${}^{1}${\it Instituto de F\'{\i}sica Te\'orica UAM/CSIC\\
C/ Nicol\'as Cabrera, 13--15,  C.U.~Cantoblanco, E-28049 Madrid, Spain}\\ 
\vspace{0.3cm}

${}^{2}${\it HEP Theory Group, Departamento de F\'{\i}sica, Universidad de
  Oviedo\\  
Avda.~Calvo Sotelo s/n, E-33007 Oviedo, Spain}\\

\vspace{1.8cm}


{\bf Abstract}

\end{center}

\begin{quotation}
  We show how \textit{colored} $\mathrm{SU}(2)$ BPS monopoles (that
  is: $\mathrm{SU}(2)$ monopoles satisfying the Bogomol'nyi equation
  whose Higgs field and magnetic charge vanish at infinity and which
  are singular at the origin) can be obtained from the BPST instanton
  by a singular dimensional reduction, explaining the origin of the
  singularity and implying that the singularity can be cured by the
  oxidation of the solution. We study the oxidation of other monopole
  solutions in this scheme.
\end{quotation}

\newpage
\pagestyle{plain}


\newpage

\section{Introduction: monopoles and instantons}
\label{sec-Kron}

It has been known for a long time that selfdual Yang--Mills (YM) instantons in
4-dimensional Euclidean space $\mathbb{E}^{4}$ and magnetic monopoles
satisfying the Bogomol'nyi equation in $\mathbb{E}^{3}$
\cite{Bogomolny:1975de}\footnote{This is the equation satisfied by the
  't~Hooft--Polyakov monopole \cite{'tHooft:1974qc,Polyakov:1974ek} in the
  Prasad-Sommerfield limit \cite{Prasad:1975kr}. We will henceforth refer to
  these monopoles as BPS monopoles. Since the time direction does not play any
  role here, we will also refer to the spatial parts of 4-dimensional
  Lorentzian solutions as ``3-dimensional'' solutions.} are related by
dimensional reduction. In its simplest setting, this relation can be described
as follows: if $\hat{A}_{\hat{\mu}}$ ($\hat{\mu}=0,1,2,3$)\footnote{We dress
  4-dimensional objects with a hat; hatless objects are 3-dimensional.} is the
gauge potential of a selfdual YM instanton solution in $\mathbb{E}^{4}$ and is
furthermore independent of one of the 4 Cartesian coordinates, $z$ say, then
the $z$-component $\hat{A}_{z}$ and the other three components $\hat{A}_{m}$
($m=1,2,3$) can be identified with the Higgs field $\Phi \equiv -\hat{A}_{z}$
and the gauge potential $A_{m} \equiv \hat{A}_{m}$ of a solution of the
Yang--Mills--Higgs (YMH) system in the Prasad-Sommerfield limit satisfying the
Bogomol'nyi equation:

\begin{equation}
\label{eq:B}
\mathcal{D}_{m}\Phi = \tfrac{1}{2}\epsilon_{mnp}F_{np}\, .
\end{equation}
 
The sign in the Bogomol'nyi equation depends on the orientation of the
coordinates; we have taken the one corresponding to $z$ to be $x^{0}$ and
$\epsilon_{0123}=\epsilon_{123}=+1$.

The coordinate $z$ has to be compactified
for the instanton action
to be finite:\footnote{This choice of period is unconventional but convenient
  for what follows.}  $z\sim z+4\pi$. Thus, in practice, we are performing the
dimensional reduction in $S^{1}\times\mathbb{E}^{3}$ and the $z$-independent
solutions can be considered to be the Fourier zero modes of instanton solutions
periodic in the direction $z$ (the so-called \textit{calorons}).

The paradigm of selfdual YM instanton in $\mathbb{E}^{4}$ is the BPST instanton
\cite{Belavin:1975fg}, usually presented in Cartesian coordinates using the
't~Hooft symbols. It belongs to a family of selfdual YM solutions depending on
an arbitrary function $K$, harmonic on $\mathbb{E}^{4}$ (see {\em e.g.\/}
Ref.~\cite{Jackiw:1976fs} and references therein). With $K$ asymptotically
constant and with a single point-like pole at the origin $K =1+4/(\lambda^{2}
\rho^{2})$, where $|\vec{x}_{(4)}|^{2}\equiv \rho^{2}$, the solution describes a
single BPST instanton located at the origin. Replacing $K$ by a harmonic function on
$S^{1}\times \mathbb{E}^{3}$ with a single pole at the origin and
asymptotically constant in $\mathbb{E}^{3}$, $K = 1 +
(\sinh{r/2})/[\lambda^{2}r^{2}(\cosh{r/2} -\cos{z/2})]$, where $\rho^{2} =
z^{2}+r^{2}= z^{2}+|\vec{x}_{(3)}|^{2}$, we get a caloron \cite{Harrington:1978ve} whose Fourier zero
mode gives, upon dimensional reduction, the spatial part of a Wu-Yang
$\mathrm{SU}(2)$ magnetic monopole \cite{Wu:1967vp}, which is singular at the
origin.

Since the BPST instanton and caloron are regular everywhere, the singularity
of the Wu--Yang solution can be understood as the result of
having ignored the massive Fourier modes in the dimensional reduction, but the
mere oxidation of the 3-dimensional monopole does not automatically restore
them: the 4-dimensional singular instanton corresponding to the Fourier zero
mode of the BPST caloron is singular.

The above \textit{redox} relation was generalized by Kronheimer in
Ref.~\cite{kn:KronheimerMScThesis} to a relation between selfdual Yang--Mills
instanton solutions in hyper--K\"ahler (HK) spaces
\cite{kn:KronheimerMScThesis} and BPS monopoles in $\mathbb{E}^{3}$. We are
going to see that Kronheimer's scheme provides an
alternative reduction of the BPST instanton which relates it to the
\textit{colored} BPS monopole solution of Protogenov
\cite{Protogenov:1977tq}. Colored monopoles are a rather misterious type of
monopole solutions that exist for many gauge groups \cite{Meessen:2015nla}
and are characterized by asymptotically vanishing Higgs field and magnetic charge which, nevertheless,
can contribute to the Bekenstein--Hawking entropy of certain (supersymmetric) non-Abelian black
holes \cite{Meessen:2008kb,Bueno:2014mea,Meessen:2015nla}.

Let us start by reviewing Kronheimer's result:  consider a
4-dimensional HK space admitting a free $\mathrm{U}(1)$ action which shifts
the adapted periodic coordinate $z\sim z+4\pi$ by an arbitrary constant. Its
metric can always be put in the form \cite{Gibbons:1979zt}

\begin{equation}
\label{eq:HK}
d\hat{s}^{\, 2} = H^{-1}(dz+\omega)^{2}+Hdx^{m}dx^{m} 
\hspace{1cm}
(m=1,2,3)\, ,
\end{equation}

\noindent
where the $z$-independent function $H$ and 1-form $\omega$ are related
by\footnote{Unhatted objects are always defined in 3-dimensional Euclidean
  space $\mathbb{E}^{3}$.}

\begin{equation}
\label{eq:dHdo}
dH = \star d\omega\, .   
\end{equation}

\noindent
The integrability condition of this equation implies that $H$ is a harmonic
function in $\mathbb{E}^{3}$ which is furthermore required to be strictly positive in order for the
metric to be regular. Now, for any gauge group G, let us consider a gauge
field $\hat{A}$ whose field strength $\hat{F}$ is selfdual $\hat{\star}
\hat{F}= +\hat{F}$ in the above HK metric with respect to the frame and
orientation

\begin{equation}
\label{eq:Kframe}
\hat{e}^{\, 0} = H^{-1/2}(dz+\omega)\, ,
\hspace{1cm}  
\hat{e}^{\, a} = H^{1/2}\delta^{a}{}_{m} dx^{m}\, ,
\hspace{1cm}  
\epsilon_{0123}=+1\, .
\end{equation}

\noindent
Then, the 3-dimensional gauge and Higgs fields $A$ and $\Phi$ defined
by

\begin{equation}
\label{eq:Kformulae}
\begin{array}{rcl}
\Phi
& \equiv &
-H\hat{A}_{z}\, ,
\\
& & \\
A_{m}
& \equiv &
\hat{A}_{m}
-\omega_{m}\hat{A}_{z}\, ,
\end{array}
\end{equation}

\noindent
satisfy the Bogomol'nyi equation  in $\mathbb{E}^{3}$ Eq.~(\ref{eq:B}). 
It is worth stressing that, had we started with an anti-selfdual YM field 
we would have obtained the Bogomol'nyi equation with opposite sign, which is
acceptable, but also Eq.~(\ref{eq:dHdo}) with opposite sign, which would be a
contradiction: in this setup we can only reduce YM fields which are
selfdual w.r.t.~the above frame and orientation. 

When $H=1$, the HK space is just $S^{1}\times\mathbb{E}^{3}$ and one recovers
the result explained at the beginning. A more interesting choice is $H=1/r$
with $r^{2}=x^{m}x^{m}$. Writing the $\mathbb{E}^{3}$ metric $dx^{m}dx^{m}$ as
$dr^{2}+r^{2}d\Omega^{2}_{(2)}$ and then redefining $r= \rho^{2}/4$ the HK
metric Eq.~(\ref{eq:HK}) becomes the metric of $\mathbb{E}^{4}$ in spherical
coordinates

\begin{equation}
\label{eq:sphericalmetric}
ds^{2} = d\rho^{2} +\rho^{2}d\Omega^{2}_{(3)}\, ,  
\end{equation}

\noindent
where $d\Omega^{2}_{(3)}$ is the round metric of the 3-sphere of unit radius
in Eq.~(\ref{eq:s3metricinEulerangleparametrization}). This HK space is,
therefore, $\mathbb{E}^{4}_{-\{0\}}$ and the shifts of $z$ act freely on it because
the origin $\rho=0$ does not belong to it.

Obviously, the standard BPST instanton is a selfdual solution in this space
and, provided that the gauge field is independent of $z$, we can reduce it
directly (avoiding the caloron step) using Kronheimer's scheme to find a
monopole in $\mathbb{E}^{3}_{-\{0\}}$. This is what we are going to do in the
next section but, before, we want to review the relation between the Euclidean
action of the instanton and the monopole charge. 

The gauge field strength components in the frame Eq.~(\ref{eq:Kframe}) are

\begin{equation}
\left\{
\begin{array}{rcl}
\hat{F}_{ab}
& = &
H^{-1}F_{ab} -H^{-2}\Phi (d\omega)_{ab}\, ,
\\
& & \\
\hat{F}_{0a}
& = &
H^{-1}\mathcal{D}_{a}\Phi  -H^{-2}\Phi \partial_{a}H\, ,
\\  
\end{array}
\right.
\end{equation}

\noindent
Substituting them into the YM action and using repeatedly Eq.~(\ref{eq:dHdo}),
the Bogomol'nyi equation (\ref{eq:B}) and Stokes' theorem we get

\begin{equation}
\tfrac{1}{4}{\displaystyle\int} d^{4}x \sqrt{|\hat{g}|}\hat{F}^{2}
 =
4\pi
 \int_{V^{3}}
\tfrac{1}{2}H^{-2} d\star dH\, \Phi^{2}
+
4\pi
\int_{\partial V^{3}}
\left[
H^{-1}\Phi^{A}F^{A} +\tfrac{1}{2} \star dH^{-1} \Phi^{2}
\right]\, ,
\end{equation}

\noindent
where $V^{3}$ is $\mathbb{E}^{3}$ with the singular points of $H$ removed:
this means that the first term on the r.h.s.~always vanishes. The end result therefore reads

\begin{equation}
\label{eq:actionformula}
\tfrac{1}{4}{\displaystyle\int} d^{4}x \sqrt{|\hat{g}|}\hat{F}^{2}
 =
4\pi
\int_{\partial V^{3}}
\left[
H^{-1}\Phi^{A}F^{A} +\tfrac{1}{2} \star dH^{-1} \Phi^{2}
\right]\, ,
\end{equation}

\noindent
and one must take into account that the boundary of $V^{3}$ includes the
singularities of $H$ as well as infinity. 

For $H=1$, $V^{3}=\mathbb{E}^{3}$ and the r.h.s.~is directly related to the
monopole magnetic charge

\begin{equation}
\label{eq:p}
p = \tfrac{1}{4\pi} \int_{S^{2}_{\infty}}
\frac{\Phi^{A}F^{A}}{\sqrt{\Phi^{B}\Phi^{B}}}\, ,  
\end{equation}

\noindent
provided the Higgs field is asymptotically constant, as in the BPS
't~Hooft--Polyakov monopole. 

For $H=1/r$, which is the case of interest here, $V^{3}=
\mathbb{E}^{3}_{-\{0\}}$, $\partial V^{3} = \{ 0 \}\cup S^{2}_{\infty}$, and
the integral will diverge precisely for monopoles with well-defined magnetic
charge at infinity and asymptotically constant Higgs fields. Thus, we can only
expect convergence for colored magnetic monopoles \cite{Meessen:2015nla}. If
the selfdual YM field has a finite action, then it must lead to a colored
monopole in $\mathbb{E}^{3}$ by Kronheimer's dimensional reduction.  In the
next section we are going to see that this is indeed the case for the BPST instanton.

\section{Singular reduction of the BPST instanton}
\label{sec-singularreduction}

In order to reduce the BPST instanton \textit{\`a la Kronheimer} in the HK
space with $H=1/r$, it is convenient to write it in spherical coordinates and,
actually, it is easier to rederive it directly using the following ansatz for
the components of the $\mathrm{SU}(2)$ gauge potential

\begin{equation}
\label{eq:ansatz1}
\hat{A}^{A}_{\begin{smallmatrix} L \\ R \\ \end{smallmatrix}} 
= 
b_{\begin{smallmatrix} L \\ R \\ \end{smallmatrix}}(\rho)
v^{A}_{\begin{smallmatrix} L \\ R \\ \end{smallmatrix}}\, ,    
\hspace{1cm}
A=1,2,3\, ,
\end{equation}

\noindent
where the $v^{A}_{\begin{smallmatrix} L \\ R \\ \end{smallmatrix}}$ are the
components of the $\mathrm{SU}(2)$ Maurer--Cartan (MC) 1-forms defined in
Eqs.~(\ref{eq:MCcomponents}), satisfying
Eq.~(\ref{eq:MCequationsincomponents}), and $b_{\begin{smallmatrix} L \\ R
    \\ \end{smallmatrix}}(\rho)$ is a function of $\rho$ to be determined by
imposing the selfduality of the gauge field strength. To this end it is most
convenient to use the frames 

\begin{equation}
\label{eq:LRframes}
\hat{e}_{\begin{smallmatrix} L \\ R \\ \end{smallmatrix}}^{\, 0} 
= 
d\rho\, ,
\hspace{1cm}
\hat{e}_{\begin{smallmatrix} L \\ R \\ \end{smallmatrix}}^{\, a} 
= 
\tfrac{1}{2}\rho\delta^{a}{}_{A} v_{\begin{smallmatrix} L \\ R \\ \end{smallmatrix}}^{A}\, ,  
\end{equation}

\noindent
for the metric Eq.~(\ref{eq:sphericalmetric}). Using the MC
1-forms it is straightforward to compute the
gauge field strength $\hat{F}_{\begin{smallmatrix} L \\ R \\ \end{smallmatrix}}^{A}$:

\begin{equation}
\hat{F}{\begin{smallmatrix} L \\ R
    \\ \end{smallmatrix}}^{A} 
= \frac{2\dot{b}}{\rho}\, \delta^{A}{}_{a}\, 
\hat{e}{\begin{smallmatrix} L \\ R  \\ \end{smallmatrix}}^{0}
\wedge 
\hat{e}{\begin{smallmatrix} L \\ R  \\ \end{smallmatrix}}^{a}
+\frac{2b(b\mp 1)}{\rho^{2}}\epsilon^{A}{}_{ab}\, 
\hat{e}{\begin{smallmatrix} L \\ R \\ \end{smallmatrix}}^{a} 
\wedge 
\hat{e}{\begin{smallmatrix} L \\ R  \\ \end{smallmatrix}}^{b}\, .  
\end{equation}

Requiring $\hat{F}_{\begin{smallmatrix} L \\ R \\ \end{smallmatrix}}^{A}$ to
be (anti-)selfdual ($\hat{F}^{A (\pm)}{}_{0a}=
\pm\tfrac{1}{2}\epsilon_{abc}\hat{F}^{A(\pm)}{}_{bc}$) in these two frames we
arrive at a differential equation for $b^{\pm}_{\begin{smallmatrix} L \\ R
    \\ \end{smallmatrix}}(\rho)$ leading to two self- and two anti-selfdual
solutions describing a single BPST instanton or anti-instanton, of size\footnote{
  In the instanton literature it is customary to denote the size of the (anti-)instanton by $\rho$, see {\em e.g.\/} Refs.~\cite{Vandoren:2008xg},
  but here we'll denote it by $\rho_{0}$. It is then easy to see that $\lambda = 2/\rho_{0}$.
}
determined by the parameter $\lambda$, at the origin:

\begin{equation}
\hat{\star}\hat{F}=+\hat{F}\,
\left\{
\begin{array}{rcl}
\hat{A}^{A(+)}_{L} 
& = & 
{\displaystyle\frac{1}{1+\lambda^{2}\rho^{2}/4}}\,  v_{L}^{A}\, , 
\\  
& & \\
\hat{A}^{A(+)}_{R} 
& = & 
-{\displaystyle\frac{\lambda^{2}\rho^{2}/4}{1+\lambda^{2} \rho^{2}/4}}\, v_{R}^{A}\, , 
\\  
\end{array}
\right.
\hspace{.4cm}
\hat{\star}\hat{F}=-\hat{F}\,  
\left\{
\begin{array}{rcl}
\hat{A}^{A(-)}_{L} 
& = & 
+{\displaystyle\frac{\lambda^{2}\rho^{2}/4}{1+\lambda^{2}\rho^{2}/4}}\, v_{L}^{A}\, , 
\\  
& & \\
\hat{A}^{A(-)}_{R} 
& = & 
-{\displaystyle\frac{1}{1+\lambda^{2} \rho^{2}/4}}\, v_{R}^{A}\, .
\\  
\end{array}
\right.
\end{equation}

The gauge fields $\hat{A}^{A(\pm)}_{L}$ are gauge-equivalent to the
$\hat{A}^{A(\pm)}_{R}$ owing to 

\begin{equation}
U \hat{A}^{A(\pm)}_{L} U^{-1} +dUU^{-1} 
 = 
\hat{A}^{A(\pm)}_{R}\, ,
\end{equation}
 
\noindent
and the property Eq.~(\ref{eq:RversusL}). Then, we could just work with
$\hat{A}^{A(+)}_{R}$ and $\hat{A}^{A(-)}_{L}$, which are regular (they vanish
at $\rho=0$ while the other two are multivalued there). However, if we want to
use Kronheimer's results we are forced to work with the singular ones,
$\hat{A}^{A(+)}_{L}$ and $\hat{A}^{A(-)}_{R}$, because as one can see
the transformation between the frame $\hat{e}_{\begin{smallmatrix} L \\
    R \\ \end{smallmatrix}}^{\hat{a}}$ in Eqs.~(\ref{eq:LRframes}) and
Kronheimer's frame $\hat{e}^{\hat{a}}$ in Eqs.~(\ref{eq:Kframe}) preserves the
orientation for $\hat{e}_{L}^{\hat{a}}$ but reverses it for
$\hat{e}_{R}^{\hat{a}}$. In other words: the regular gauge fields
$\hat{A}^{A(+)}_{R}$ and $\hat{A}^{A(-)}_{L}$ are anti-selfdual in
Kronheimer's frame and can therfore not be consistently reduced.

Let us, then, consider $\hat{A}^{A(+)}_{L}$ and $\hat{A}^{A(-)}_{R}$. By
construction, these gauge fields are invariant under the free $\mathrm{U}(1)$
actions in Eqs.~(\ref{eq:leftu1action}) and (\ref{eq:rightu1action}),
respectively.

In other words: $\hat{A}^{A(+)}_{L}$ is $\varphi$-independent and
$\hat{A}^{A(-)}_{R}$ is $\psi$-independent  and can  be
dimensionally reduced along those directions because the only invariant point
under these actions (the origin $\rho=0$) does not belong to our HK space. We
can expect 3-dimensional monopoles  which are singular there.

Using directly Eqs.~(\ref{eq:Kformulae}), from $\hat{A}^{A(+)}_{L}$ we get the
Yang--Mills and Higgs fields of a BPS monopole solution

\begin{equation}
\label{eq:coloured}
\Phi^{A(+)}_{L}
= 
\frac{1}{r(1+\lambda^{2}r)}\delta^{A}{}_{m} \frac{y_{L}^{m}}{r}\, ,  
\hspace{1cm}
A^{A(+)}_{L}
= 
\frac{1}{(1+\lambda^{2}r)} \epsilon^{A}{}_{mn}d \frac{y_{L}^{m}}{r} \frac{y_{L}^{n}}{r}
\end{equation}

\noindent
where we have defined the Cartesian coordinates $y^{m}/r \equiv
-\delta^{m}{}_{A}v^{A}_{L\, \varphi}$:\footnote{We use the identity
  $v^{A}_{L}(\varphi=0)-\cos{\theta}\, v^{A}_{L\, \varphi} d\psi =
  \epsilon^{A}{}_{mn}d \frac{y_{L}^{m}}{r} \frac{y_{L}^{n}}{r}$}

\begin{equation}
y_{L}^{1}\equiv  r\sin{\theta}\, \cos{\psi}\, ,
\hspace{.5cm}
y_{L}^{2}\equiv  r\sin{\theta}\, \sin{\psi}\, , 
\hspace{.5cm}
y_{L}^{3}\equiv  r\cos{\theta}\, .  
\end{equation}

The reduction of $\hat{A}^{A(-)}_{R}$ gives exactly the same 3-dimensional
fields upon the replacement of the Cartesian coordinates $y^{m}_{L}$ by
$y^{m}_{R} \equiv +r\delta^{m}{}_{A}v^{A}_{R\, \psi}$:\footnote{Now we use the identity
  $v^{A}_{R}(\psi=0)-\cos{\theta}\, v^{A}_{R\, \psi} d\varphi =
  -\epsilon^{A}{}_{mn}d \frac{y_{R}^{m}}{r} \frac{y_{R}^{n}}{r}$}

\begin{equation}
y_{R}^{1}\equiv  r\sin{\theta}\, \cos{\varphi}\, ,
\hspace{.5cm}
y_{R}^{2}\equiv  -r\sin{\theta}\, \sin{\varphi}\, , 
\hspace{.5cm}
y_{R}^{3}\equiv  -r\cos{\theta}\, .  
\end{equation}

As predicted by the arguments based on the Euclidean action, the 3-dimensional
BPS monopole obtained by this procedure is the colored monopole found by
Protogenov in Ref.~\cite{Protogenov:1977tq}. The Higgs field vanishes at
infinity and the magnetic charge, as defined in Eq.~(\ref{eq:p}) vanishes
identically. The solution approaches the Wu--Yang monopole \cite{Wu:1967vp}
for $r\rightarrow 0$ (which corresponds to $\lambda^{2}=0$) and, therefore,
one can argue that the solution describes a magnetic monopole at the origin
whose charge is completely screened at infinity. This interpretation is
supported by the computation of the Bekenstein--Hawking entropy $S_{\rm BH}$
of non-Abelian black holes with this kind of gauge fields: there is a
contribution to $S_{\rm BH}$ corresponding to a magnetic charge
\cite{Meessen:2008kb,Bueno:2014mea}. 

\section{Oxidation of the singular Protogenov monopoles}

Reversing the procedure we just carried out, we see that the
singularity of the $\mathrm{SU}(2)$ colored BPS monopole disappears completely
when it is oxidized to 4 Euclidean dimensions. Since there are other singular
$\mathrm{SU}(2)$ BPS monopoles \cite{Protogenov:1977tq}, it is natural to ask
whether their singularities can also be cured by oxidizing them within this scheme.

The spherically symmetric solutions of the $\mathrm{SU}(2)$ Bogomol'nyi
equations have the following \textit{hedgehog} form \cite{Protogenov:1977tq}:

\begin{eqnarray}
A^{A} 
& = &
-r^2 h(r) \epsilon^{A}{}_{mn} \frac{y^{n}}{r} d \left( \frac{y^{m}}{r} \right) \, ,
\\
& & \nonumber \\
\Phi^{A} 
& = & 
-r f(r) \delta^{A}{}_{m} \frac{y^{m}}{r} \, ,
\end{eqnarray}

\noindent
where the functions $f(r)$ and $h(r)$ must satisfy the differential equations 

\begin{eqnarray}
\label{eq:effe}
r \dot{h} +2h+f (1+r^{2} h ) 
& = & 
0 \, , 
\\
& & \nonumber \\
\label{eq:effeb}
r  (\dot{h}-\dot{f})-r^{2}h(h-f) 
& = & 0 \, ,
\end{eqnarray}

\noindent
if the above Yang-Mills and Higgs fields are to satisfy the Bogomol'nyi equation
(\ref{eq:B}). Apart from the family of colored solutions in
Eq.~(\ref{eq:coloured}), there is another 2-parameter ($\mu$ and $s$) family
of solutions given by

\begin{equation}
\label{eq:Protogenovequations}
rf 
 =  
-\frac{1}{r}
\left[
1-\mu r\coth{(\mu r+s)} 
\right]
\, ,
\hspace{1cm}
rh
 = 
\frac{1}{r}
\left[
\frac{\mu r}{\sinh{(\mu r+s)}} 
-1
\right]
\, .
\end{equation}

The BPS limit of the 't~Hooft--Polyakov monopole
\cite{'tHooft:1974qc,Polyakov:1974ek} is the $s=0$ member of this family, and
the only regular one. Before oxidizing them, we can compute the action of the
corresponding instanton using Eq.~(\ref{eq:actionformula}). The action turns
out to diverge for all values of $s$. However, even if all hope of getting a
regular instanton 
by oxidizing these solutions is lost, it is still worth
finding the general expression of the singular instantons, since it may give
us inspiration for making instanton ans\"atze directly in 4 dimensions. Using
Kronheimer's relations, Eq.~(\ref{eq:Kformulae}), we find 

\begin{equation}
\hat{A}^{A} = -r^{2} f(r) v_{L}^{A} + r^{2} \left[ f(r)-h(r) \right] u^{A}\, ,
\end{equation}

\noindent
where we have defined the 1-forms 

\begin{equation}
  \begin{array}{rcl}
u^{1} 
& = & 
\cos{\psi} \sin{\theta} \cos{\theta} d\psi+\sin{\psi} d\theta  \, , 
\\[2pt]
u^{2} 
& = & 
\sin{\psi} \sin{\theta} \cos\theta d\psi -\cos{\psi} d\theta  \, , 
\\[2pt]
u^{3} 
& = & 
-\sin^{2}{\theta} d\psi \, .\\
\end{array}
\end{equation}

\noindent
These 1-forms depend only on two coordinates ($\psi$ and $\theta$) and they
can be seen as projections of the left-invariant MC 1-forms $v_{L}^{A}$

\begin{equation}
u^{A}= v_{L}^{B}\left[\delta^{A}{}_{B} -\frac{y_{B}y^{A}}{r^{2}} \right]\, .
\end{equation}

\noindent
They satisfy differential equations identical to the ones satisfied by the
left-invariant MC 1-forms $v_{L}^{A}$ up to the $1/2$ factor, {\em i.e.\/}

\begin{equation}
du^{A}=-\epsilon^{A}{}_{BC} u^{B} \wedge u^{C}\, ,
\end{equation}

\noindent
which makes them well suited for a generalization of the ansatz
Eq.~(\ref{eq:ansatz1}):

\begin{equation}
\hat{A}^{A} = b(\rho) v^{A}_{L}+c(\rho)u^{A}\, .  
\end{equation}

\noindent
Imposing selfduality of the corresponding field strength with the
redefinition 

\begin{eqnarray}
b(\rho(r))
= 
-r^{2} f(r)\, ,
\hspace{1cm}
c(\rho(r))
=
-r^{2} \left[h(r)-f(r) \right]\, ,
\end{eqnarray}

\noindent
leads to Protogenov's equations (\ref{eq:effe}) and (\ref{eq:effeb}); the oxidation
of the BPS monopoles gives all the selfdual instantons of that form.

\section{Conclusions}
\label{sec-conclusions}

In this paper we have shown how a misterious kind of $\mathrm{SU}(2)$ BPS
magnetic monopoles known as \textit{colored monopoles}, which are singular at
the origin and have vanishing asymptotic charge and Higgs field, can be
understood as the result of the singular dimensional reduction of the BPST
instanton, which is itself globally regular.  The parameter appearing in the
monopole family of solutions turns out to be related to the one that measures
the instantons' size.

The mechanism is analogous to the well-known mechanism curing gravitational
singularities by oxidation as for example the KK-monopole \cite{Sorkin:1983ns}
or in certain 4-dimensional dilatonic black holes \cite{Gibbons:1994vm}, but
with the twist that here the fields are non-Abelian.  The mechanism that cures
the singularity of the colored monopole does not, however, work for the rest
of the spherically-symmetric BPS monopoles of the theory: they always have
infinite action, but depending on the application this may or may not be a
problem.

We have argued, based on the relation between the instanton action and the
monopole magnetic charge, that this relation between regular instantons and
singular, colored magnetic monopoles should be general. It has recently been
shown in Ref.~\cite{Meessen:2015nla} that colored magnetic monopoles are
present in the Yang--Mills--Higgs theory for all $\mathrm{SU}(N)$ groups and
the results of that paper can be used to construct regular selfdual
$\mathrm{SU}(N)$ instantons \cite{kn:MOR}. Possibly, the transmutation
monopoles discovered in Ref.~\cite{Meessen:2015nla}, which have different
(non-vanishing) charges at infinity and at the origin, can be related to
regular solutions by a similar mechanism.

The case studied here is just the simplest and most special of those comprised
in Kronheimer's work Ref.~\cite{kn:KronheimerMScThesis}, since it just
involves $\mathbb{E}^{4}_{-\{0\}}$. One may wonder if the rest can be of any
relevance in physics. It turns out that the relation between $\mathcal{N}=1,
d=5$ and $\mathcal{N}=2, d=4$ super-Einstein--Yang--Mills (SEYM) theories must
include the relation between selfdual instantons in HK spaces and BPS
monopoles in $\mathbb{E}^{3}$ discovered by Kronheimer: the timelike
supersymmetric solutions of $\mathcal{N}=1, d=5$ \cite{Bellorin:2007yp} (as
it happens in the Abelian case \cite{Gauntlett:2002nw}) involve a
4-dimensional Euclidean base space of HK type and the YM field strengths have
a piece which is selfdual in that space.  On the other hand the YM fields of
the timelike supersymmetric solutions of $\mathcal{N}=2, d=4$ SEYM
\cite{Meessen:2012sr} are required to satisfy the Bogomol'nyi equation in
$\mathbb{E}^{3}$ in combination with an effective \textit{Higgs} field. These
two classes of theories and their solutions are related by dimensional
reduction. Explicit solutions of the latter describing non-Abelian black holes
have been obtained in
\cite{Huebscher:2007hj,Huebscher:2008yz,Meessen:2008kb,Bueno:2014mea,Meessen:2015nla}. Some
of the solutions are powered by the colored BPS monopoles that we have shown
to be related to the BPST instanton. It is then natural to expect that the
oxidation of the complete supergravity solutions will provide us with explicit
solutions of the $\mathcal{N}=1, d=5$ SEYM theory\footnote{So far, no
  explicit solutions of these theories have been constructed.} involving the
BPST instanton. These solutions, whose form is quite intriguing, may be
globally regular. The oxidation \textit{\`a la Kronheimer} of solutions
involving other monopoles will give potentially singular solutions, but, just
as it happens with singular monopoles in $d=4$, gravity may cover the
singularities with event horizons. All these new possibilities opened by the
result presented in this paper are very interesting and well worth
investigating. Work in this direction is already under way \cite{kn:MOR2}.

\section*{Acknowledgments}

PM wishes to thank D.~Rodr\'{\i}guez-G\'omez for stimulating discussions.
This work has been supported in part by the Spanish Ministry of Science and
Education grants FPA2012-35043-C02 (-01 {\&} -02),
the Centro de Excelencia Severo Ochoa Program grant SEV-2012-0249, 
the EU-COST Action MP1210 ``The String Theory Universe'',
the Principado de Asturias grant GRUPIN14-108
and the Spanish Consolider-Ingenio 2010 program CPAN CSD2007-00042.
The work was further supported by the JAE-predoc grant JAEPre 2011 00452 (PB)
and the \textit{Severo Ochoa} pre-doctoral grant SVP-2013-067903 (PF-R). 
T.O. would like to thank M.M.~Fern\'andez for her permanent support. 

\appendix

\section{The metrics of the round $\mathrm{S}^{3}$ and $\mathrm{S}^{2}$}
\label{sec-S3S2}

In this appendix we will review the well-known construction of the
$\mathrm{SO}(4)$-invariant metric on $\mathrm{S}^{3}$ using its identification
with the $\mathrm{SU}(2)$ group manifold, the construction of
$\mathrm{SO}(3)$-invariant metric on $\mathrm{S}^{2}$ using its identification
with the $\mathrm{SU}(2)/\mathrm{U}(1)$ coset space and the relation between
both of them.

All matrices $U\in \mathrm{SU}(2)$ ($U^{\dagger}=U^{-1}$, $\mathrm{det}\,
U=+1$) can be parametrized by two complex numbers $z_{0},z_{1}$

\begin{equation}
\label{eq:Udef}
U\equiv 
\left(
 \begin{array}{rr}
z_{0} & z_{1} \\
-\bar{z}_{1} & \bar{z}_{0} \\
 \end{array}
\right)\!, 
\hspace{1.5cm}
|z_{0}|^{2}+|z_{1}|^{2}=1\, .
\end{equation}

\noindent
Therefore, the $\mathrm{SU}(2)$ manifold can be identified with
$\mathrm{S}^{3}$. Both are traditionally parametrized by the Euler angles
$\{\theta,\varphi,\psi \}$:

\begin{equation}
z_{0}= \cos\!\left(\theta/2\right)\, e^{i(\varphi+\psi)/2}\, ,
\hspace{1cm} 
z_{1}= \sin\!\left(\theta/2\right)\, e^{i(\varphi-\psi)/2}\, .
\end{equation}

\noindent
The main property of this parametrization is that any $\mathrm{SU}(2)$
rotation can be written as the product of three rotations with these angles:

\begin{equation}
U(\varphi,\theta,\psi)
= 
U(\varphi,0,0)U(0,\theta,0) U(0,0,\psi)\, .
\end{equation}

The Euler angles are usually assumed to take values in the intervals $\theta
\in [0,\pi]$, $\varphi \in [0,2\pi)$, and $\psi\in [0,4\pi)$. Other choices
are possible: for instance, $\theta \in [0,\pi]$, $\varphi \in [0,4\pi)$, and
$\psi\in [0,2\pi)$ also covers once $\mathrm{S}^{3}$. Only the coordinate
chosen to take values in $[0,4\pi)$ should be considered periodic. There is a
free $\mathrm{U}(1)$ action on $\mathrm{S}^{3}$ associated to constant shifts
of the periodic coordinate. For the standard choice, this action is

\begin{equation}
\label{eq:rightu1action}
U(\varphi,\theta,\psi)
\rightarrow
U(\varphi,\theta,\psi)
U(0,0,2\alpha)\, ,
\hspace{1cm}
\alpha \in [0,2\pi)\, .
\end{equation}

\noindent
Being a right action, it is adequate to define the right coset space
$\mathrm{SU}(2)/\mathrm{U}(1)$. If we choose instead $\varphi$ to be the
periodic coordinate, the $\mathrm{U}(1)$ action is

\begin{equation}
\label{eq:leftu1action}
U(\varphi,\theta,\psi)
\rightarrow
U(2\alpha,0,0)U(\varphi,\theta,\psi)\, ,
\hspace{1cm}
\alpha \in [0,2\pi)\, .
\end{equation}

\noindent
Being a left action, it is adequate to define the left coset space
$\mathrm{U}(1)\backslash \mathrm{SU}(2)$, which is a more unusual option.

A convenient basis of the $\mathfrak{su}(2)$  Lie algebra is provided by the
anti-Hermitian matrices\footnote{The $\sigma^{A}$ are the Pauli matrices,
which we take to satisfy
\begin{equation}
\sigma^{A}\sigma^{B} = \delta^{AB} +i\epsilon^{ABC}\sigma^{C}\, .
\end{equation}
}

\begin{equation}
\label{eq:su2Tigenerators2}
T_{A}=\tfrac{i}{2}\sigma^{A}\, ,
\hspace{1cm}
[T_{A},T_{A}] = -\epsilon_{ABC}T_{C}\, .
\end{equation}

\noindent
In this basis

\begin{equation}
U(\varphi,0,0) = e^{\varphi T_{3}}\, ,
\hspace{1cm}
U(0,\theta,0) = e^{\theta T_{2}}\, ,
\hspace{1cm}
U(0,0,\psi) = e^{\psi T_{3}}\, .  
\end{equation}

The left- (resp.~right-)invariant Maurer--Cartan (MC) 1-form $V_{ L}$
(resp.~$V_{R}$) are defined by

\begin{equation}
\label{eq:MCdef}
V_{L}\equiv -U^{-1}dU \, ,
\hspace{1cm}
V_{R}\equiv -dUU^{-1}\, ,
\end{equation}

\noindent
and as a consequence of their definition they satisfy the MC equations

\begin{equation}
dV_{\begin{smallmatrix} L \\ R \\ \end{smallmatrix}}  
\mp
V_{\begin{smallmatrix} L \\ R \\ \end{smallmatrix}} 
\wedge 
V_{\begin{smallmatrix} L \\ R \\ \end{smallmatrix}}
=
0\, .
\end{equation}

Observe that the left- and right-invariant MC 1-forms are related
by the following \textit{gauge} transformations:

\begin{equation}
\label{eq:RversusL}
V_{R} = UV_{L}U^{-1}\, .  
\end{equation}

The components of the MC 1-forms in the above basis $V_{\begin{smallmatrix} L
    \\ R \\ \end{smallmatrix}} \equiv v_{\begin{smallmatrix} L \\ R
    \\ \end{smallmatrix}}^{A} T_{A}$ are given by

\begin{equation}
\label{eq:MCcomponents}
\left\{
 \begin{array}{rcl}
v_{L}^{1} 
& = & 
\sin\psi\, d\theta -\sin\theta \cos\psi\, d\varphi\, ,
\\[2pt]
v_{L}^{2} 
& = & 
-\cos\psi\, d\theta -\sin\theta \sin\psi\, d\varphi\, ,
\\[2pt]
v_{L}^{3} 
& = & 
-(d\psi +\cos\theta\, d\varphi)\, ,
\\
 \end{array}
\right.
\hspace{1cm}
\left\{
 \begin{array}{rcl}
v_{R}^{1} 
& = & 
-\sin\varphi\, d\theta +\sin\theta \cos\varphi\, d\psi\, ,
\\[2pt]
v_{R}^{2} 
& = & 
-\cos\varphi\, d\theta -\sin\theta \sin\varphi\, d\psi\, ,
\\[2pt]
v_{R}^{3} 
& = & 
-(d\varphi +\cos\theta\, d\psi)\, ,\\
 \end{array}
\right.
\end{equation}

\noindent
and the MC equations in components take the form

\begin{equation}
\label{eq:MCequationsincomponents}
dv_{\begin{smallmatrix} L \\ R \\ \end{smallmatrix}}^{A}
\pm 
\tfrac{1}{2}\epsilon_{ABC}\,
v_{\begin{smallmatrix} L \\ R \\ \end{smallmatrix}}^{B}
\wedge 
v_{\begin{smallmatrix} L \\ R \\ \end{smallmatrix}}^{C}
=
0\, .
\end{equation}

As their name indicates, the left- (resp.~right-)invariant MC 1-forms are
invariant under the left (resp.~right) $\mathrm{U}(1)$ action in
Eq.~(\ref{eq:leftu1action}) (resp.~Eq.~(\ref{eq:rightu1action})).

Both the left- or the right-invariant MC 1-forms can be used as
Dreibeins to construct a bi-invariant (that is $\mathrm{SU}(2)\times
\mathrm{SU}(2)\sim \mathrm{SO}(4)$ -invariant) metric on $\mathrm{SU}(2)$
($\sim\mathrm{S}^{3}$) with tangent space metric $\delta_{AB}$. The result is
exactly the same in both cases: normalizing the metric so as to get the volume
of the 3-sphere of unit radius, we find

\begin{equation}
\label{eq:s3metricinEulerangleparametrization}
d\Omega^{2}_{(3)}
=
\tfrac{1}{4} v_{L}^{A}v_{L}^{A} 
=
\tfrac{1}{4} v_{R}^{A}v_{R}^{A}
= 
\tfrac{1}{4}\!
\left[
d\theta^{2} +d\varphi^{2} +d\psi^{2} +2\cos\theta \, d\varphi d\psi
\right]\, .
\end{equation}

\noindent
It is customary to rewrite this metric so that the invariance  under the chosen
$\mathrm{U}(1)$ action is manifest. For the standard choice in which
$\psi \in [0,4\pi)$ is the periodic coordinate and there is invariance under
the right action in Eq.~(\ref{eq:rightu1action}) 

\begin{equation}
d\Omega^{2}_{(3)}
=  
\tfrac{1}{4}
\left[
d\Omega^{2}_{(2)}(\theta,\varphi)
+v_{L}^{3}v_{L}^{3}
\right]\, ,
\end{equation}

\noindent
where $d\Omega^{2}_{(2)}(\theta,\varphi)$ is the standard metric of
the round 2-sphere of unit radius

\begin{equation}
\label{eq:metrics2}
d\Omega^{2}_{(2)}(\theta,\varphi)  
= 
d\theta^{2} + \sin^{2}\theta d\varphi^{2}
=
v_{L}^{1}v_{L}^{1} +v_{L}^{2}v_{L}^{2}\, .
\end{equation}

\noindent
For the other choice, we just have to interchange $\varphi$ and $\psi$ and $L$
by $R$ in the above expressions.


\end{document}